\documentstyle[11pt,aaspp4]{article}
\font\sb=cmr10
\def\CIV{C$\,${\sb IV}$\,\lambda$1550}
\def\MSUN{${\rm M}_\odot$}
\def\kms{km sec$^{-1}$} 
\begin{document}

\title{Optical-IR  Spectral Energy Distribution of the Proto-Galaxy 
Candidate MS1512-cB58}

\author{
E.~Ellingson}
\affil{CASA, CB 389, University of Colorado, Boulder, CO 80309 \\
Email: e.elling@casa.colorado.edu}
\author{H.~K.~C.~Yee\altaffilmark{1}}
\affil{Department of Astronomy, University of Toronto, Toronto,
Ontario M5S 3H8, Canada, \\ and \\
Canada-France-Hawaii Telescope, P.~O.~Box 1597, Kamuela,
Hi 96743 \\
Email: hyee@astro.utoronto.ca}
\author{Jill Bechtold\altaffilmark{2}}
\affil{Steward Observatory, University of Arizona, Tucson AZ 85721
\\ Email: jbechtold@as.arizona.edu}
\author{R. Elston\altaffilmark{2}}
\affil{National Optical Astronomy Observatories, CTIO, Casilla
603, La Serena, Chile 1353.\altaffilmark{3} \\
Email: elston@ctio.noao.edu
}
\altaffiltext{1}{Guest observer, Canada-France-Hawaii Telescope,
operated jointly by the NRC of Canada, CNRS of France, and the
University of Hawaii}
\altaffiltext{2}{Visiting Astronomer at the Infrared Telescope
Facility which is operated by the University of Hawaii under
contract to the National Aeronautics and Space Administration}
\altaffiltext{3}{Operated by the Association of Universities
for Research in Astronomy, Inc, under contract with the National
Aeronautics and Space Administration}

\received{}
\accepted{}

\begin{abstract}
The spectral energy distribution of 
the proto-galaxy candidate MS1512-cB58
at $z =2.72$  discovered by Yee et al. (1996) is presented. Photometry
in seven bands ranging from $g$ to $K$$'$ (1300--6000 \AA~rest wavelength) are
fitted with population synthesis models from Bruzual \& Charlot
(1993). The data confirm a very young age for this galaxy, in  agreement
with ages estimated from preliminary \CIV~P-Cygni profile modeling.
Single-burst models with ages greater than about 20 Myr can be discarded
at the 99\% confidence level, and
continuous star formation models with ages greater than about 35 Myr
can be discarded at the 95\% confidence level.
The spectral energy distribution is most consistent with
a continuous star formation
model of about 10--20 Myr, with reddening of $E(B-V)\sim 0.3$. 
No evidence for an older population of stars
is seen, but the  possibility of
an older population with as much as 90\% of the galaxy mass 
cannot be ruled out. We discuss the possible
ramifications of a non-standard IMF and gravitational
lensing on the galaxy's age and mass. 

\end{abstract}

\noindent

\section{INTRODUCTION}

The identification of the high-redshift precursors of normal
present-epoch galaxies, and specifically a galaxy
in its first episode of star formation, has long been an elusive goal
(e.g., Partridge \& Peebles 1967).
In past decades, a wide variety of models have been
suggested and techniques attempted to identify
such proto-galaxies (see Koo 1986 and Pritchet 1994 for reviews). 
The proto-galaxy candidate serendipitously discovered by
Yee et al. (1996; hereafter Paper I) may offer an excellent chance to study
a very young galaxy in detail.  The $V=20.64$ galaxy, designated as
MS1512-cB58, (hereafter, cB58), lies in the field of 
the $z=0.37$ galaxy cluster MS1512+36, which was  observed  as part of
the Canadian Network
for Observational Cosmology (CNOC) cluster redshift survey
(see Yee, Ellingson \& Carlberg 1996). 
The galaxy was found to have $z=2.72$ by identification of about
a dozen strong absorption lines indicative of young stars.
Optical colors--- Gunn $g$, $V$, Gunn $r$ and Johnson $I$--- indicated
that the stellar population of this galaxy is 
less than 400 Myr old.
However, the uncertain extinction
correction made it difficult to constrain the galaxy
age to better than a range of 10--400 Myr.
P-Cygni profiles in the \CIV~absorption
lines suggested that the rest UV flux is dominated by a population
as young as 10 Myr old. 

In this paper we present data in three additional photometric bands in the
infrared: $J$, $H$ and $K$$'$. These data allow stronger limits to be placed
on both the extinction and the age of the stellar populations
in this galaxy. We also explore the possibility of underlying
older populations and discuss briefly the effects of
non-standard  initial mass functions and gravitational
lensing on our conclusions.

\section{Observations and Data}

$J$ and $K$$'$ images were obtained on  1995 Sep 28 UT
at the NASA Infrared-Telescope Facility (IRTF) with the NSFCAM
imager, which contains a 256x256 InSb array (Shure \& Rayner 1993).
The plate scale was 0.30$''$ per pixel.
The data were obtained under photometric conditions with
seeing 0.8 -- 0.9$''$.
For $K$$'$, a 25 point grid 
with integration times of 15 seconds, 4 co-adds each, was used, resulting
in a  total integration time of 1500 seconds.  For $J$, a 15 point
grid was used with 20 second exposures, 3 co-adds each, resulting in 
900 seconds of integration.
The UKIRT Faint Standards \#28 and 26 (Casali \& Hawarden 1992)
were observed for calibration just before and just after 
the galaxy observations.
The data were reduced using 
standard techniques.

Images in the $H$ band were obtained at the CFHT 3.6m using
the Redeye-Wide Infrared Camera with a 256$\times$256 pixel NICMOS3
HgCdTe array on the night of 1995 Dec 30 UT, under photometric skies.
The pixel size of the detector is 0.5$''$ per pixel. 
Six sets of 9 dithered frames of 35 seconds exposure 
were obtained, providing a total integration time of 1890 seconds.
The photometry was calibrated using observations of the UKIRT Faint Standard
Star \#23 (Casali \& Hawarden 1992) taken
before and after the galaxy observations. 

Photometry in the optical bands of $g$, $V$, $r$, and $I$
is taken from Paper I.
The photometry for the near IR observations was derived in
the same way as the optical data (see Paper I), using an
aperture of 2.2$''$ diameter for the object and 9$''$ for the
outer diameter of the sky aperture.
The 2.2$''$ aperture photometry is then corrected to the
``total'' magnitude at an aperture
equivalent to the $V$ isophote of 24.1 mag arcsec$^{-2}$,
assuming zero color gradients.
The photometric data along with their uncertainties
are listed in Table 1.
Also listed are the magnitudes in the AB system,
where corrections for the IR photometry
were based on the 0 mag fluxes from Wamsteker (1981).

\section{Discussion}

\subsection{Spectral Energy Distribution Models}

In order to test the calibration of photometry gathered from
several different sources, the spectral energy distributions (SEDs) of
galaxies of known redshift near cB58 are first examined. 
Figure 1 plots the data for the  cD galaxy of the cluster 
and a bright companion galaxy located 6$''$ W and 18$''$ S of 
the cD (designated \#100983 in the CNOC catalog, Abraham 
et al. 1996).
The cD galaxy 
has a spectrum of an old population of stars but with strong
[OII] emission, and possibly a somewhat blue continuum--
both presumably associated with the cooling flow
detected in this cluster (Donahue, Stocke \& Gioia 1992). 
Figure 1 also shows GISSEL
spectral synthesis models from Bruzual \& Charlot (1993).
Both models are for a 13 Gyr old single burst (coeval) population with 
a Salpeter initial mass function (IMF). 
This standard IMF has an upper mass cutoff
125\MSUN, and a lower mass cutoff of 0.1\MSUN.
The models fit the data reasonably well,
indicating that there are probably no gross inconsistencies 
in the calibration of the SEDs. Note that the model is slightly
lower than the cD data bluewards of the 4000 \AA ~break, as expected. 

In Paper I, it was pointed out that interpreting the age of
a stellar population from rest UV data is complicated by 
dust extinction. 
The SED was modeled assuming extinction  occurs in the rest frame,
and a slightly modified version of the average LMC curve from
Fitzpatrick (1986) was used. 
Both here and in Paper I, the 
2175\AA~hump in the LMC extinction was interpolated over, 
based on the fact that there is no evidence for the expected
prominent dip due to this feature in the optical spectroscopy data.
Note that many local star burst galaxies also do not
show this bump in the extinction in their IUE
spectra (Kinney et al.~1993). 
An extinction law derived from
UV observations of starburst galaxies (Calzetti et al. 1994)
 was also tested and found to yield qualitatively the same
 results. 
 The extinction in this case 
 has a significantly grayer slope than that 
 from the LMC, and hence requires a larger amount of extinction
 to provide enough  reddening to fit the SED.
This in turn will increase the already high
 rest optical luminosity of this galaxy. 
 Hence, the more conservative LMC law is adopted.

Three single burst GISSEL models were presented in Paper I to
illustrate the range of parameters allowed by the optical
data: 400 Myr old with no extinction,
200 Myr old with the equivalent of $E(B-V)=0.18$, and 10 Myr old with
$E(B-V)=0.3$. 
The oldest single burst model
possible was 400 Myr, and that the maximum dust extinction 
possible was about $E(B-V)\sim0.4$, since the intrinsic 
spectrum would become
bluer than the hottest stars if more dust is assumed.
In Figure 2, these
three models are plotted again, this time including the IR data.     
The spectral range is now from 1300--6000 \AA~rest.
Each model is normalized to match the optical data, as in Paper I, and 
assumes a Salpeter IMF with the mass limits noted above.
It is clear that the only model which comes close to fitting
the new IR data is the
youngest model, of a 10 Myr old single burst and $E(B-V)=0.3$.
The older models clearly predict too much IR flux in the $J$ through $K'$
bands to match the data. No single burst model fits
the data within the 95\% confidence level and we can conservatively
rule out all models older than 20 Myr at the 99\% confidence level.
Note that the $J$ and $H$ bands straddle the 4000 \AA~
and Balmer breaks and can be used alone as an age indicator which is
relatively insensitive to extinction. The blue slope
of the spectrum in this region clearly indicates that the
SED is dominated by a young population. 

Continuous star formation models provide a more physically plausible
scenario. 
For these models, the rest UV flux is always dominated by very young stars, 
and in Paper I it  was found that no age discrimination 
was possible using just the optical data. However, over time 
it is expected that the observed IR flux will increase.
Figure 3 shows continuous star formation models of
ages 5, 10, 50, and 500 Myr, assuming
a Salpeter IMF.  Here the data have been corrected
for extinction equivalent to $E(B-V)=0.3$ to match
the observed optical data. The IR data clearly favor the 10 Myr
continuous star formation model, and renormalization
of the models relative to the data also provides good
fits to ages as high as 20 Myr. Models older than
about 35 Myr are ruled out at the 95\% confidence level.
Both of these estimates are consistent with preliminary
 modeling in Paper I of the
P-Cygni profile of the \CIV~absorption line, which suggests
a lower limit of 10 Myr for a continuous star formation
scenario.

\subsection{Limits On a Possible Older Population} 

The IR data show that the observed episode of star formation
is relatively young, but does it represent the very
first incidence of star formation in this galaxy?
Limits on the size of an underlying older population
can be estimated from these data but 
are relatively broad, 
because even at $K'$ (rest $V$), the continuum light is still
dominated by the young population. 
The SED is modeled as two episodes of
star formation: the observed burst, which is modeled as a 10 Myr
old continuous episode of star formation (referred to hereafter as 
the 10Myr model), and
an older population, which is modeled as a single
burst model with varying age and mass relative to the 10Myr model. 
Both models assume the standard Salpeter IMF.
For all combinations, the best fits to the data suggest 
no contribution from an older population.
Upper limits at the 95\% confidence level were determined for the
fraction of total stellar mass in the older population.
Note that as the population ages, its SED
differs more in shape
from the 10Myr model. However, the rapid fading of the 
single burst SED
with age makes it possible to hide a slightly increasing mass
fraction underneath the younger distribution.
Thus, the constraints on the mass fraction of the
older population are a weak function of its age,
ranging from 83\% of the total stellar mass for a
100 Myr old population, to 90\% for a 2 Gyr old population. 
Figure 3 also shows an example of a 1 Gyr old single burst
combined with the 10Myr model,
where the older population has a mass equal to 85\% of the 
combined stellar mass. This model can be ruled out
at the 95\% confidence level.
The high redshift of cB58 indicates
an upper limit on the age of any older population
of about  2 Gyr, depending on the cosmological parameters assumed.
Thus, it is possible to conclude  that there is no
evidence for an older stellar population, and
most conservatively that
cB58 has formed at least 10\% of its stars within the
previous 10-20 Myr.

\subsection{Total Stellar Mass and Star Formation Rates}
 
A lower limit on the total stellar mass can be estimated by  
noting that the $K'$ band is approximately equivalent to rest
$V$ at $z=2.72$. The $K'_{AB}$ magnitude of 19.61 therefore
implies an extinction-corrected rest $V$ absolute magnitude of --27.94 mag
($H_{0}$ = 75 km s$^{-1}$ Mpc$^{-1}$, $q_0$=0.1, A$_V$=0.93 mag). 
Assuming a 10 Myr old single burst model with no underlying older population,
the galaxy is expected to
fade approximately 6 mag over the next 10 Gyr (the lookback
time to $z=2.72$).
Note that after 10 Gyr the differences
between an instantaneous burst and just 10 Myr of continuous star
formation would be extremely subtle, so the choice of
the initial model is not important.
Thus, the observed star formation in cB58 would be the precursor 
of a galaxy at least as bright as --21.9 mag, or about 2.7 $L^*$.
Any subsequent star formation or the presence of
an additional older population would serve to make
its present-epoch luminosity even  greater, which
argues against a significant older population.

The high luminosity of cB58  implies prodigious rates of
star formation. 
Paper I estimates 
star formation  rates of 4700 \MSUN~yr$^{-1}$
based on the 1500\AA~flux and continuous star
formation models by Leitherer,
Robert, \& Heckman (1995).
The homogeneous appearance of the galaxy  suggests that
this intense  star formation is not localized, but
must be occurring across the 30 kpc visible galaxy
on close to dynamical timescales.
At this rate, a $L^*$ galaxy would be created in
a few 10's of Myr. 
It is not clear whether such a short,
intense phase of star formation over the entire galaxy is
dynamically possible.
It is also unclear whether 
the implied supernova rates would allow the galaxy to
retain any gas for further star formation, or whether
supernova-driven winds would quickly  strip the galaxy of
its interstellar medium.

The implied star formation rates  and total stellar  masses are  
highly dependent on the
initial mass function assumed. 
Variations in
the upper mass cutoff do not affect the age of
the young population, although they can change the star
formation rate by up to a factor of about 5 (Paper I, Leitherer et al. 1995).
Lower-mass cutoffs as high as 10\MSUN~have
been predicted for regions of intense
star formation (e.g. Scalo 1990).
Evidence for such a cutoff has been mixed for nearby
star forming galaxies, with some galaxies
showing near-normal IMFs (e.g. Hunter et al.~1995)
and others showing indications of a lower-mass cutoff
(e.g. Rieke et al.~1993).
The available GISSEL models only allow fits to models
up to a fairly small 2.5\MSUN~cutoff.
Such a low-mass cutoff provides a slightly poorer fit to the
observed SED, but is not
significantly worse and cannot be ruled out.
More extreme cutoffs would lower the IR flux of the models,
but since massive stars contribute most of
the SED, even into the IR, the data are not sufficient
to discern between a slightly younger
population and a more extreme cutoff.
Thus, a lower mass cutoff at masses greater than
2.5\MSUN~might allow a somewhat older age for the galaxy.
A similar conclusion would be reached for a flatter IMF.
A low-mass cutoff would of course also significantly lower the
inferred star formation rate and total stellar mass for
cB58.
Episodes of intense star formation with an extreme low-mass cutoff would
temporarily increase the galaxy's luminosity without adding significantly
to its stellar mass.
In this case, the example of cB58 would not add greatly to
our understanding of the origin of present-day stellar populations
in normal galaxies.  

The object is clearly extremely different from many other
proto-galaxy candidates in its stellar SED and high
luminosity. In comparison with the galaxies at $z \sim 3$ studied
by Steidel et al.~(1996), cB58 is about 40 times brighter
(assuming the same extinction),
and may have slightly higher equivalent width 
stellar absorption lines. An important difference is
that the $(R-K)_{AB}$ color of cB58 is 
0.8 mag, bluer than the average value of 1.3 mag for the
Steidel et al.~sample. 
This may indicate more dust in
the fainter galaxies or that they have an
older stellar population on average. If the color difference is
from dust, an additional $E(B-V)$ of 0.1--0.2 (for 
the LMC and Calzetti et al.~extinction laws, respectively)  
in the Steidel et al.~galaxies is necessary to explain the 
color difference. In this case, cB58 would be still a factor
of 20 brighter in the rest $V$ band.
Ages for the Steidel et al. galaxies of greater than 100 Myr 
would also explain the color difference. 
If continuous
star formation for 10 and 100 Myr is assumed for cB58 and
the Steidel et al.~sample, respectively, then cB58 is undergoing
star formation at a rate of about 40--100 times that of the Steidel 
et al.~sample. 

An important possibility to consider is that
the high luminosity of cB58 is in part due to gravitational lensing
from the foreground cluster MS1512+36. The galaxy is located
only 6$''$ from the cluster cD and is a very strong candidate
for lensing. However, the image is resolved on both major and minor
axes, appears quite  homogeneous and its surface brightness profile 
can even be fit with an exponential disk model (Paper I).
Williams \& Lewis (1996) have modeled the possible
lensing of cB58, assuming a subcritical cluster potential.
They found that if
the cluster mass of E1512+36 is well represented by its
velocity dispersion (690 \kms; Carlberg et al. 1996),
and Abell richness (class 0; Abraham et al. 1996),
then the lensing magnification is at most about 2 magnitudes.
However, if the cluster mass is a factor of 2 larger, 
and assuming a specific geometry,
then up to a factor of 40 magnification can be obtained
without obvious distortion in the ground-based images.
This factor of 2 in mass seems difficult to come by, as it requires
a 3--$\sigma$ error on the velocity dispersion determination
(Carlberg et al.~1996).
Williams \& Lewis note
that the X-ray luminosity of the cluster is higher than the
velocity dispersion suggests, but the presence of a cooling flow
in the cluster will tend to increase the observed X-ray luminosity. 
Furthermore, the X-ray luminosity and the measured velocity
dispersion of MS1512+36 are well within
the scatter of correlation of these quantities found by  
Edge \& Stewart (1991).
Thus
there seems to be only a small possibility that this object
is highly lensed, although  magnification by  factors of a few 
would not be surprising.

Unless the gravitational lensing magnification is much larger
than the cluster dynamics and image morphology suggest, 
the star formation rates in this galaxy are probably on
the order of at least several hundred to a thousand solar masses per year
and are distributed fairly homogeneously
across the entire galaxy.  Clearly such rates cannot
persist for long, and
thus we expect a subsequent decrease in star formation and
luminosity of this galaxy in future times.
This initial high rate of star formation and later decline 
suggests that cB58 may be a precursor of a 1--3$L^*$ early-type galaxy.
If the lensing magnification is much larger, then the stellar
mass of cB58 would be correspondingly smaller, but its 
young age and the fraction of stars in the observed 
10--20 Myr old episode of star formation would be unchanged. 
In this case, cB58 may represent the initial stages
of any of a wide variety of galaxy types.

\acknowledgements
\leftline{ACKNOWLEDGMENTS}
We thank the staff of the IRTF, 
especially the telescope operator David Griep, for 
making these observations possible.
The CFHT observations were made possible by a generous
grant of director's discretionary time by Pierre Couturier.
E.E. would like to thank Peter Conti for numerous useful discussions.
H.Y. wishes to thank CFHT for their hospitality while
this work was being done.
H.Y. is supported by an operating grant from NSERC of Canada.

\newpage
\begin{deluxetable}{ccccc}
\tablenum{1}
\tablewidth{0pt}
\tablecaption{Photometry (3$''$ aperture)}
\tablehead{
\colhead{Band}           &
\colhead{Central $\lambda$ (\AA)}           &
\colhead{Magnitude}    & \colhead{$\pm$\phm{~~~~}}  &
\colhead{AB Magnitude}    }
\startdata
 $g$ & $\,$~4930      & 21.08 & 0.10\phm{~~~~}  & 21.15 \nl
 $V$ & $\,$~5500      & 20.64 & 0.12\phm{~~~~}  & 20.64 \nl
 $r$ & $\,$~6540      & 20.60 & 0.10\phm{~~~~}  & 20.41 \nl
 $I$ & $\,$~8060      & 19.92 & 0.12\phm{~~~~}  & 20.35 \nl
 $J$ & 12400     & 19.12 & 0.13\phm{~~~~}  & 19.95 \nl
 $H$ & 16000     & 18.42 & 0.10\phm{~~~~}  & 19.82 \nl
 $K'$ & 21400     & 17.83 & 0.12\phm{~~~~}  & 19.61 \nl
\enddata
\end{deluxetable}

\newpage

\figcaption{ Optical-IR spectral energy distribution for
a) the cluster cD at z=0.37 and b) a probable cluster galaxy.
Wavelengths are rest wavelengths.
}

\figcaption{ Comparison of single burst star formation
models with the extinction-corrected  spectral energy distribution.
The three models represent (from top to bottom),
a 10 Myr old single burst with $E(B-V)=0.3$, a
200 Myr old burst with $E(B-V)=0.18$ and a 
400 Myr old burst with $E(B-V)=0$.
All models are scaled to match the optical data
(1200--3000 \AA~rest) and the dashed lines connect
the SED points with the same extinction correction.
The 10 Myr model and data are shifted by +1 in the log
for clarity.
}

\figcaption{ Comparison of continuous star formation
models with the observed SED. The data are corrected for
extinction equivalent to $E(B-V)=0.3$. The three
solid lines represent (from bottom to top): 5, 10, 50 and
500 Myr old models. The dashed line is
a combination the 10Myr model (15\% by mass)
and a 1 Gyr old single burst model (85\% by mass). 
The models are scaled to 
match the optical data (1200--3000 \AA~rest).
}

\begin{figure}[h] \figurenum{1}\plotone{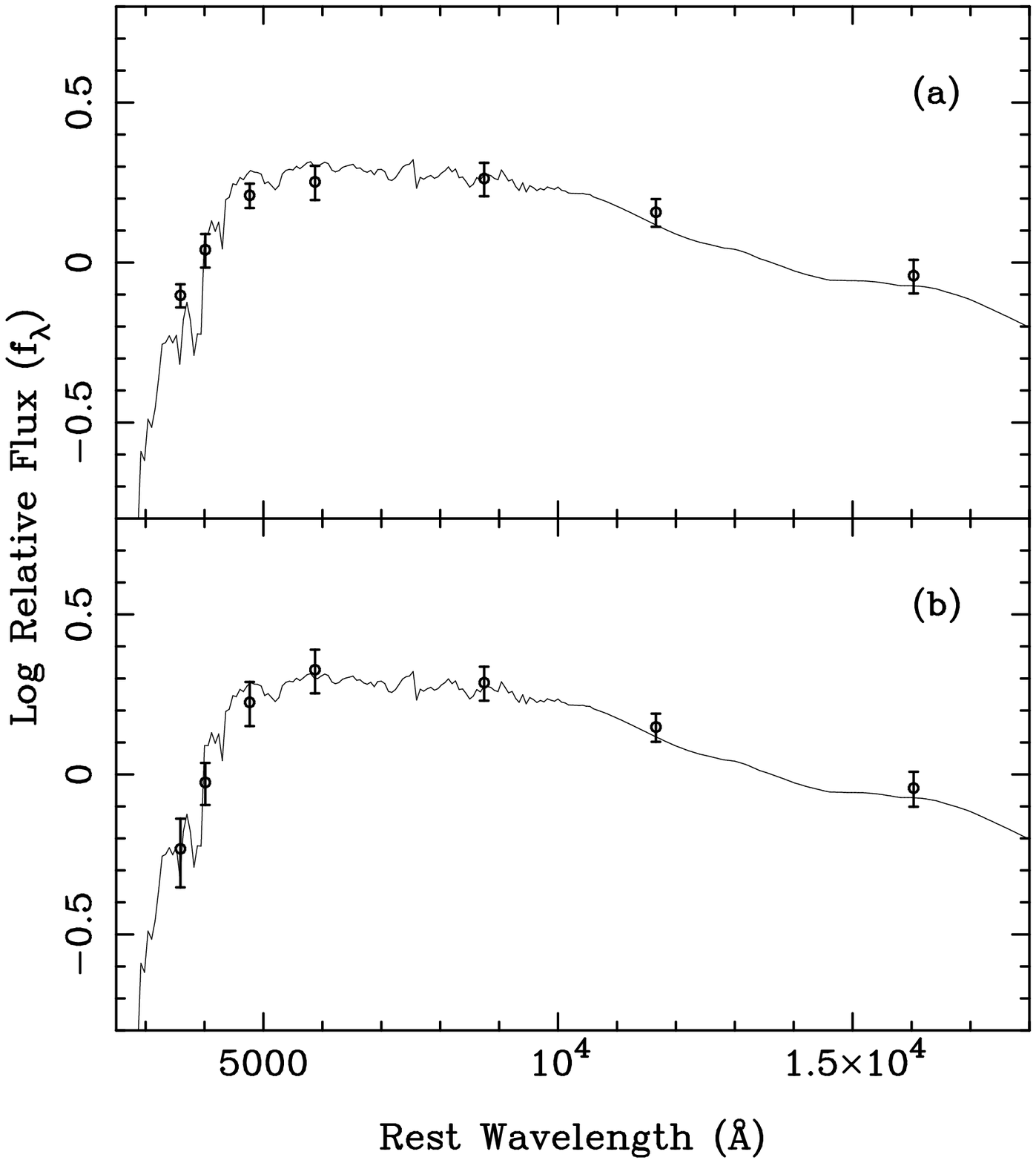} \caption{}\end{figure}
\begin{figure}[h] \figurenum{2}\plotone{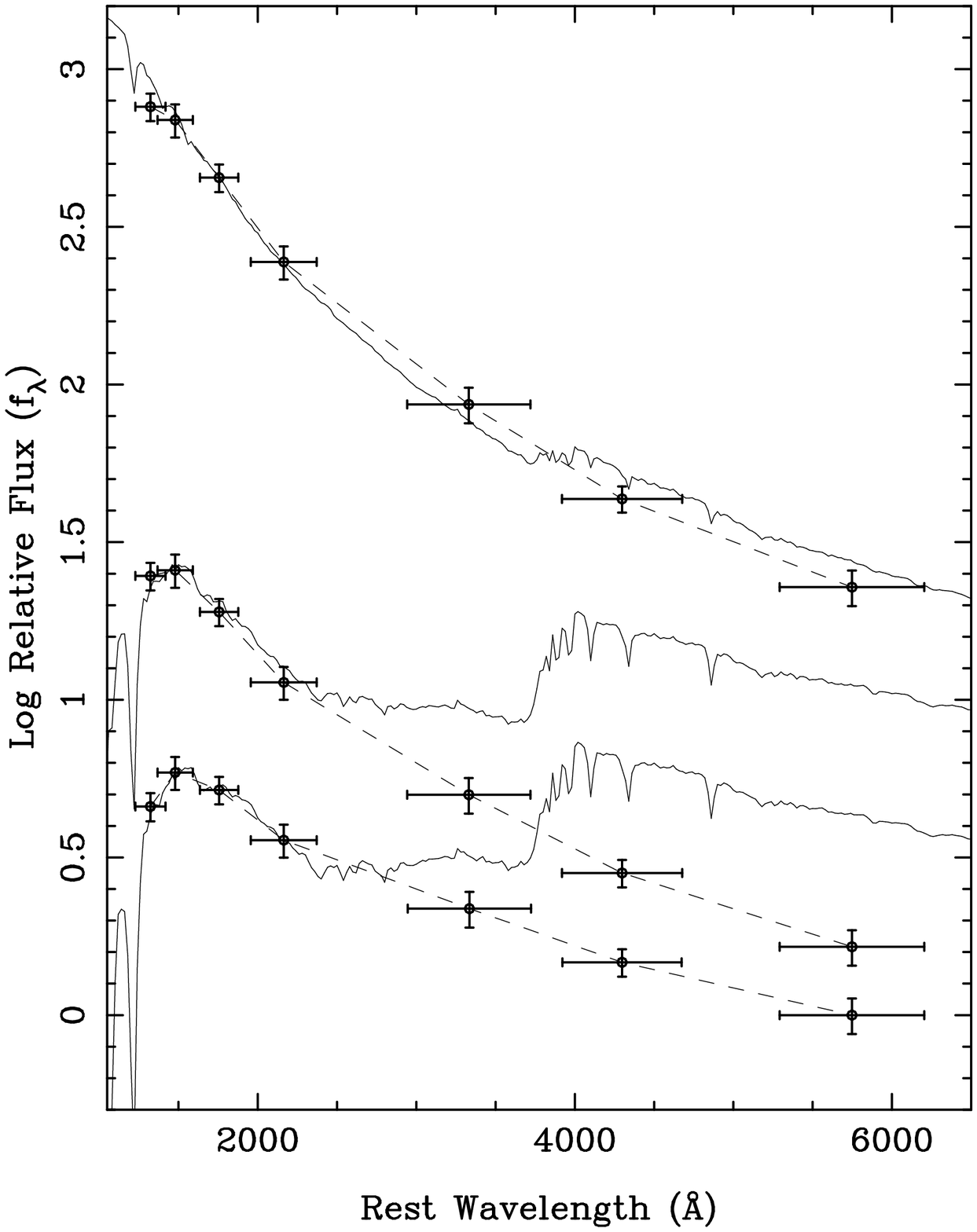} \caption{}\end{figure}
\begin{figure}[h] \figurenum{3}\plotone{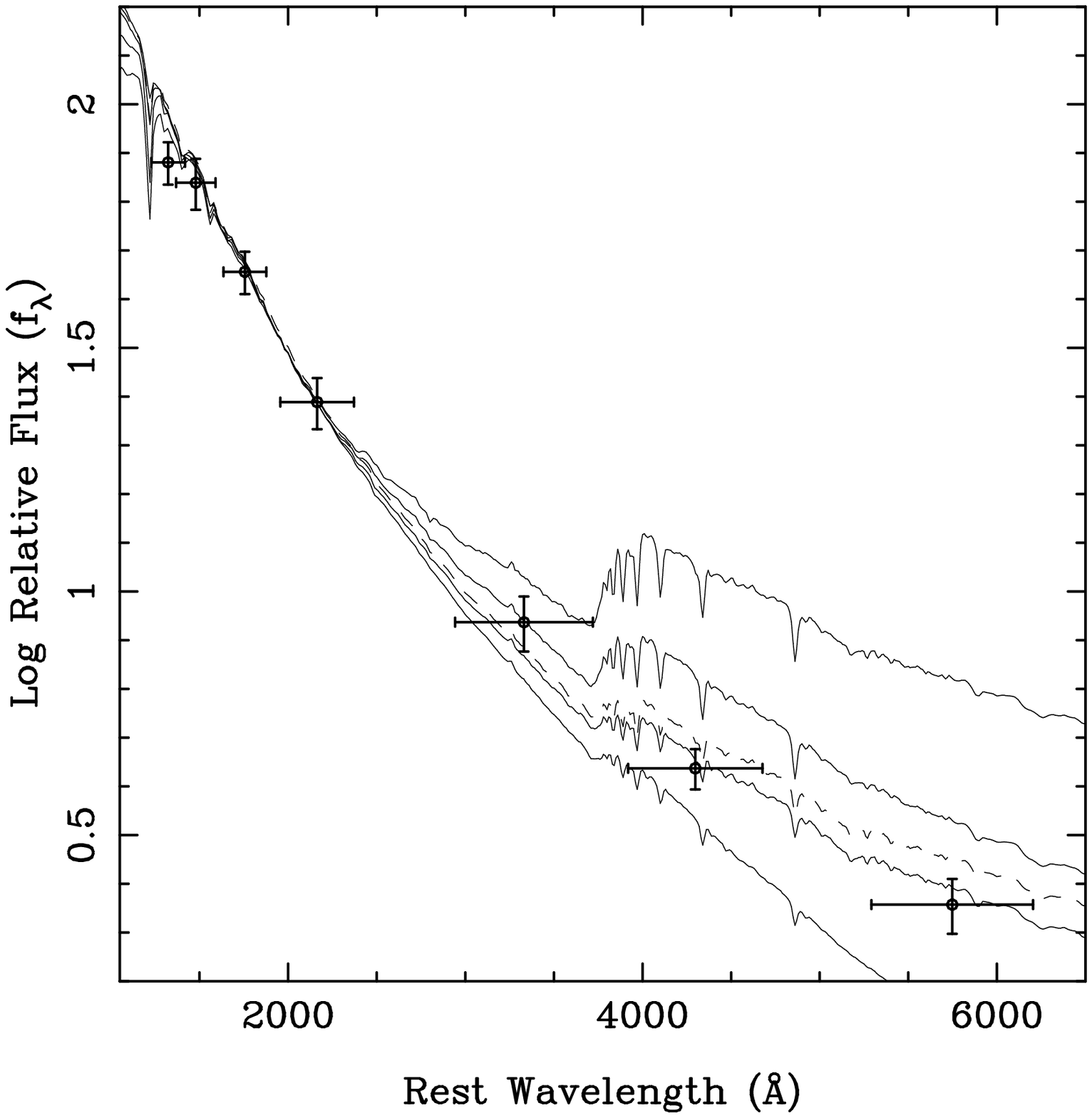} \caption{}\end{figure}

\end{document}